\documentstyle[epsf]{article}
\renewcommand{\theequation}{\arabic{equation}}

\begin{document}

\renewcommand{\theequation}{\arabic{equation}}

\begin{center}
{\Large {\bf Quantum Spin Chains with Nonlocally-Correlated Random
Exchange Coupling and Random-Mass Dirac Fermions}}

\vskip 1cm
{\Large Koujin Takeda\footnote{e-mail address:
takeda@icrr.u-tokyo.ac.jp}}
\vskip 0.5cm
{Institute for Cosmic Ray Research, University of Tokyo, Kashiwa,
Chiba, 277-8582 Japan}

\vskip 0.5cm

{\Large  Ikuo Ichinose\footnote{e-mail 
 address: ikuo@ks.kyy.nitech.ac.jp}}  
\vskip 0.5cm
{ Department of Electrical and Computer Engineering,
Nagoya Institute of Technology, Gokiso, Showa-ku,
Nagoya, 466-8555 Japan}

\end{center}

\begin{center} 
\begin{bf}
Abstract
\end{bf}
\end{center}
$S={1\over 2}$ quantum spin chains and ladders with random 
exchange coupling are studied
by using an effective low-energy field theory and transfer
matrix methods.
Effects of the nonlocal correlations of exchange couplings
are investigated numerically.
In particular we calculate localization length of magnons, density
of states, correlation functions and multifractal exponents
as a function of the 
correlation length of the exchange couplings.
As the correlation length increases, there occurs a ``phase
transition" and the above quantities exhibit different behaviors
in two phases.
This suggests that the strong-randomness fixed point of the
random spin chains and random-singlet state get unstable
by the long-range correlations of the random
exchange couplings.

\newpage

\section{Introduction}

The properties of low-energy excitations in disordered systems in 
low dimensions have been intensively studied in recent several years.
It is believed that in one-dimensional (1D) systems almost all states are
exponentially localized regardless of amount of disorder if the 
random variables
have $\delta$-function like short-range correlations.
However recently, existence of phases of extended states was reported
for some 1D models, which include the Anderson model, the aperiodic 
Kronig-Penny model, etc., if the disorder has long-range 
correlations\cite{extended}.
In the previous papers\cite{TTIK,KI} motivated by the above studies, 
we studied a field-theory model, the random-mass
Dirac fermions in 1D, which describe low-energy excitations of the
random-hopping tight-binding model.
In particular we considered the effects of the power-law spatial correlations
of the random Dirac mass 
and found that properties of the states near the band center change
as the correlation is increased, though extended states exist only
at the band center.

In this paper we apply the previous field-theory model to the
systems of spin chains and ladders with random exchange coupling.
In particular, we shall numerically calculate the localization length (LL)
of the magnons, their density of states (DOS) and correlations
of low-energy wave functions as a function of the correlation length
of the random exchange couplings.
We shall employ the Fourier filtering method (FFM) and its modified
techniques for generating random variables with long-range correlations.
It is known that in the random quantum spin chain with $S={1\over 2}$
there exists a nontrivial low-energy fixed point called strong-randomness
fixed point and at that fixed point random-singlet(RS) phase is 
realized\cite{DF}.
In particular there the spin correlation function behaves as
$\langle S^z_nS^z_0\rangle \propto {1\over n^2}$ where $n$ is the 
lattice site index.
Stability of the fixed point has been discussed from various
aspects\cite{stability}.
In this paper we study effects of the long-range correlation of the
exchange couplings of the spin chains on the stability
problem.
In order to investigate this problem, we focus on low-energy modes
and study a more tractable field-theory model, i.e., random-mass
Dirac fermions.

In Sec.2, we shall briefly review the derivation of the effective
field-theory model, i.e., the random-mass Dirac fermions, for
low-energy excitations in $S={1\over 2}$ quantum spin chains and ladders.
We specify the random exchange couplings and random mass with nonlocal
correlations.
In Sec.3, we explain FFM and modified FFM (MFFM) and show the random
variables generated by the MFFM actually have power-law correlations.
This result is very important for the present study because
the original FFM is {\em not} suitable for generating random numbers with
long-range correlations.
In Sec.4, we calculate the LL and DOS for
various random variables with nonlocal correlations by the MFFM.
These quantities change their behavior as the correlation of the
random variables is varied.
This result agrees with our previous calculations which employed the FFM and 
the above conclusion about the ``existence of phase transition" is 
confirmed though
its validity was questioned sometimes.
In Sec.5, we shall study properties of the low-energy wave
functions and the multifractal exponents.
These quantities also change their behavior for the various correlations
of the random mass.
This result strongly indicates that the long-range correlation of the
exchange couplings changes behavior of the spin-spin correlation
functions from that of the RS phase.
Section 6 is devoted for conclusion.


\section{Random spin chains and random-mass Dirac fermions}

Let us start with the random XX spin chain whose Hamiltonian is given as
\begin{eqnarray}
H_{XX}&=&-\sum_n {J_n \over 2}\Big(S^x_nS^x_{n+1}
+S^y_nS^y_{n+1}\Big) \nonumber \\
&=&-\sum_n {J_n \over 2}\Big(S^+_nS^-_{n+1}
+S^-_nS^+_{n+1}\Big),
\label{HXX}
\end{eqnarray}
where $n$ is site index and the magnitude of the quantum spin 
$S^i_n\; (i=x,y,z)$ is $1/2$.
The exchange couplings $J_n$ are random variables which have a spatial
nontrivial correlation as we specify shortly.
It is well-known that the Hamiltonian (\ref{HXX}) is mapped to a fermion
system through the Jordan-Wigner transformation,
\begin{equation}
S^z_n=C^\dagger_nC_n-{1 \over 2}, \;\; S^+_n=C^\dagger_n\exp(i\pi 
\sum^{n-1}_{i=-\infty}N_i),
\label{JW}
\end{equation}
where $C_i$ is a spinless fermion operator and $N_i=C^\dagger_iC_i$,
\begin{equation}
H_{XX}=-\sum_n {J_n \over 2}\Big(C^\dagger_nC_{n+1}+C^\dagger_{n+1}C_n\Big).
\label{HXX2}
\end{equation}
In the case of the non-random XX model with a constant exchange coupling 
$J_n=J$, the ground state of the Hamiltonian (\ref{HXX2}) is the half-filled 
state of the fermion $C_n$ and low-energy excitations are described by the
right and left-moving modes near the Fermi points $k_F=\pm {\pi\over 2}$,
\begin{equation}
C_n=(i)^n\psi_R(n)+(-i)^n\psi_L(n).
\label{RL}
\end{equation}
In this paper we consider the random exchange couplings $J_n$
which have a uniform constant part $J$ and a random fluctuating
part $\delta J_n=(-)^n m(n)$, where $m(n)$ is a ``smooth function"
of site index $n$;
\begin{eqnarray}
J_n&=&J+\delta J_n  \nonumber \\
&=& J+(-)^n m(n).
\label{Jn}
\end{eqnarray}
By substituting Eqs.(\ref{RL}) and (\ref{Jn}) into (\ref{HXX2}),
we obtain the low-energy effective field theory of the random
XX model,
\begin{eqnarray}
{\cal H}&=&\int dx \psi^\dagger h\psi, \nonumber   \\ 
h&=&-i\sigma^z \partial_x +m(x)\sigma^y,
\label{Hdirac}
\end{eqnarray}
where $\vec{\sigma}$ are the Pauli spin matrices, $m(x)$ is the
``continuum limit" of $m(n)$ ($x=na$ with the lattice spacing $a$),
$\psi(x)=(\psi_R(x),\psi_L(x))^t$
and we have put $J=1$ without loss of
generality.
The above Hamiltonian is nothing but the random-mass Dirac field
in 1D which we studied in the previous papers\cite{TTIK,KI}.
The LL and DOS were calculated by the transfer-matrix methods
and imaginary vector potential methods.
>From (\ref{Hdirac}), it is obvious that
an energy gap appears for the constant nonvanishing mass $m(x)=m_0$.
This corresponds to the energy gap in the dimer 
or spin-Peierls state of the XX
spin chain which appears as a result of the alternating exchange couplings 
$J_n=J+(-)^nm_0$.
There the spin-spin correlation function (SSCF) 
$[\langle S^z_nS^z_0 \rangle]_{\rm ens}$
decays exponentially as $e^{-m_0|n|}$ ($[\;\; ]_{\rm ens}$ means the
ensemble average for the random variables).
We are interested in how this spin-Peierls state changes when $J_n$'s become
random variables. 

It is not so difficult to show that the following random
Heisenberg spin chain has the same low-energy effective theory
with the random XX spin chain\cite{SFG},
\begin{equation}
H=\sum_nJ_n\vec{S}_n\cdot \vec{S}_{n+1}.
\end{equation}
This result comes from the fact that in the spin-Peierls state for 
$J_n=J+(-)^nm_0$ the term $S^z_nS^z_{n+1}$ becomes irrelevant because
of the existence of the spin gap.
In the uniform case $J_n=J$ (i.e., $m_0=0$), 
the above term is marginal and renormalize
the ``speed of light" and the exponent of the spin 
correlation\cite{affleck}.
In the {\em random} spin chains on the other hands, 
it is believed that the RS phase dominates at low energies
and the SSCF behaves as 
$$
[\langle S^z_nS^z_0 \rangle]_{\rm ens} \propto {1\over n^2}
$$
regardless of the (relatively small) value of the $z$ component of spin 
coupling.
It can be also shown that the spin ladder system like
\begin{equation}
H_{SL}=\sum_{n, j=1,2}J\vec{S}_{n,j}\cdot\vec{S}_{n+1,j}
+\sum_n J_{\bot n}\vec{S}_{n,1}\cdot\vec{S}_{n,2}
\end{equation}
has a low-energy excitations which is described by the random-mass
Dirac fermions\cite{SFG,SNT}.
There the Dirac mass is proportional to the inter-ladder coupling
$J_{\bot n}$.

We assume a nontrivial spatial correlation for the random Dirac mass $m(x)$
in (\ref{Hdirac}),
\begin{equation}
  [\ m(x) \;m(y)  \ ]_{\rm ens}=  \chi{(|x-y|)}.
\label{disordercor}
\end{equation}
For the short-range white-noise correlation,
which is often employed in studies of the random systems,
$\chi(|x-y|)\propto \delta(x-y)$.
In this paper we are interested in the effects of nonlocal
correlations, i.e., the exponential-decay correlation like
\begin{equation}
[\ m(x) \; m(y)\ ]_{\rm ens} = \frac{g}{2 \lambda} \; 
\exp (\frac{-|x-y|}{\lambda}),
\label{expC}
\end{equation}
and also the power-decay correlation like,
\begin{equation}
[\ m(x) \; m(y)\ ]_{\rm ens}={C \over |x-y|^{\gamma}},
\label{powerC}
\end{equation}
where $g, \; \lambda, \; C$ and $\gamma$ are parameters.

We shall consider various nonlocally-correlated random exchange couplings
or random Dirac masses and calculate the LL of the magnon, DOS and SSCF.
Behaviors of these quantities change as the correlation of the random 
mass is varied.
As in the previous papers\cite{KI}, we shall consider multi-kink configurations
of the mass $m(x)$.
In the practical calculation, we fix the distances between kinks
and vary the magnitudes of $m(x)$ as random variables (see Fig.1).
In the previous papers, we used the FFM for generating random numbers with
nontrivial correlation.
The FFM has some disadvantage for large systems.
To avoid this, we first generated a large sequence of random numbers and
used only small fraction of them for practical calculations of the LL and DOS.
In this paper we shall employ a modified method of the FFM 
which was proposed by Makse et.al.\cite{Makse}.
Let us first explain the MFFM briefly.

\begin{figure}
\label{fig:example}
\begin{center}
\unitlength=1cm

\begin{picture}(17,3)
\centerline{
\epsfysize=3.5cm
\epsfbox{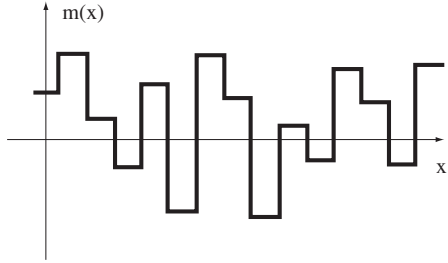}
}
\end{picture}
\caption{\small An example of random m(x).}
\end{center}
\end{figure}


\section{Long-range Correlated Disorders}

We explain the methods for generating long-range correlated random
variables numerically in this section.
The FFM is a well-known numerical method for
generating random numbers with correlations. 
However, this method is not suitable for generating power-law
correlated random numbers. Random numbers generated by this method 
 are power-law correlated within only 0.1\% of the whole system.
The MFFM was proposed by Makse et.al.\cite{Makse} in order to
modify this flaw. 
They showed the random numbers generated by the MFFM have the
 power-law correlation almost over the whole system.

Here we briefly review the FFM and the MFFM.
First we consider the random numbers $u_{i}$ which have {\em white-noise}
correlation,
$[ u_{i} u_{i+l} ]_{\rm ens} = \delta_{l,0}$.
Our goal is to generate random numbers $\{\eta_{i}\}$ which has the power-law 
correlation like,
 \begin{equation}
 \label{ffm1}
 C(l)=[ \eta_{i} \eta_{i+l} ]_{\rm ens} \sim l^{-\gamma}
 \end{equation}
 where $\gamma$ is the exponent of the power decay.
It is straightforward to show that
variables $\eta_q$, the Fourier transformation of $\eta_i$ in Eq.(\ref{ffm1}),
also has the power-law correlation with respect to $q$ as follows,
\begin{equation}
 \label{ffm2} 
 S(q)=[ \eta_{q} \eta_{-q} ]_{\rm ens} \sim q^{\gamma-1}.
\end{equation}
Then one can easily show that Eq.(\ref{ffm2}) is satisfied if we put,
 \begin{equation}   
 \label{ffm3}
 \eta_{q} = \{S(q)\}^{1/2} u_{q}
 \end{equation}
 where $u_{q}$ is the Fourier transformation of $u_{i}$.

>From the above observation, the FFM is given as follows; 
 First we generate white-noise random numbers $u_{i}$ and
 calculate its Fourier transformation $u_{q}$. 
Secondly we calculate $\eta_{q}$ using
 Eq.(\ref{ffm3}). Thirdly we calculate the inverse Fourier transformation of 
 $\eta_{q}$ and obtain the power-law correlated random numbers
 $\eta_{i}$.

 The modified FFM changes the following part of the original FFM.
 The power-law correlation $C(l)$ has the singularity at the origin
$l=0$. This singularity causes the difficulty of large-distance
correlation which we mentioned above.
 Therefore we replace it as follows,
 \begin{equation}
 \label{ffm4}
 C(l) = (1+l^{2})^{-\gamma /2}
 \end{equation}
 which has the same power-law correlation with Eq.(\ref{ffm1}) for large $l$.
 We also impose periodic boundary condition on $C(l)$,
 \begin{equation}
 C(l) = C(l+L)
 \end{equation}
 where $L$ is system size, and we define $C(l)$ by Eq.(\ref{ffm4}) only 
in the range 
 $ -L/2 < l < L/2 $. 
In practical calculations using the MFFM,
 we obtain $S(q)$ from $C(l)$ in Eq.(\ref{ffm4}) by using numerical
 Fourier transformation.   
 
 We show the two-point correlation function of random numbers $\eta_{i}$
 generated by the MFFM in Fig.2. 
 The averaged correlation functions show power-law behavior
 almost exactly over half size of the system.  
This result is much better than those obtained by the old-fashioned FFM.
In this paper we apply the MFFM to generate the random exchange 
coupling of the quantum 
spin model or random mass of Dirac fermion with power-law correlations.
The correlator $[ m(x)m(y) ]_{\rm ens}$ has the same behavior with
$[ \eta_i\eta_{i+l} ]_{\rm ens}$ in Fig.2.
>From this result it is obvious that $[ m(x)m(y) ]_{\rm ens}$
has the desired behavior in almost the whole system.

\begin{figure}
\label{fig:correlation}
\begin{center}
\unitlength=1cm

\begin{picture}(17,3)
\centerline{
\epsfysize=5cm
\epsfbox{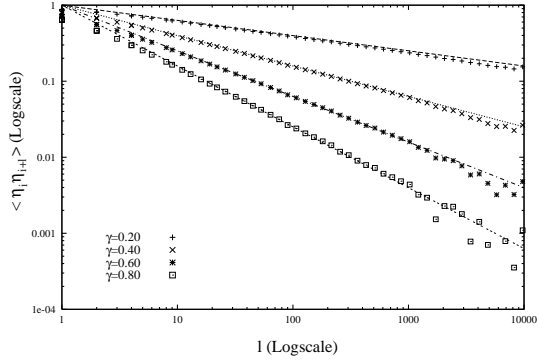}
}
\end{picture}
\caption{\small The correlation function of $\eta_{i}$ generated by modified
 FFM: $\eta_{i}$ is designed to have power-law correlation.
 We set number of sites $2^{14}=16384$ here. $\gamma$ is defined in 
 Eq.(\ref{ffm1}).}
\end{center}
\end{figure}

\section{Localization Length and DOS}

In this section we show the result of numerical calculations of 
the LL and DOS in the 
system with the nonlocally correlated random exchange
couplings or random-mass Dirac fermions.
By using the {\em modified} FFM, we first calculate the LL
and DOS as a function of
energy and reexamine the previous results obtained by
the {\em old-fashioned} FFM.

 We calculate energy dependence of the averaged (typical) localization length
 of eigenstates and DOS first.
 In one-dimensional disorder system, the
 DOS and localization length are related by the Thouless
 formula\cite{Thouless}.
 In the case of random mass Dirac fermion (or random spin 
 chain), this formula is given as,
\begin{equation}
\label{eq:thouless}
\frac{1}{\xi(E)} \propto \int_{0}^{|E|} \log(\frac{E}{E'}) \rho(E') dE',
\end{equation} 
 where $\xi(E)$ and $\rho(E)$ are the LL and DOS
 at energy $E$, respectively.
 It is not so difficult to show that the DOS is parametrized as
\cite{IK}
 \begin{equation}
 \rho(E)\sim E^{-\eta}|\log E|^{-\delta},
 \label{rhoE}
 \end{equation}
 where $\eta$ and $\delta$ are exponents.
Then the Thouless formula 
 Eq.(\ref{eq:thouless}) gives the energy dependence of
the LL as follows,
 \begin{eqnarray}
 \xi(E) &\sim& E^{-1+\eta}|\log E|^{-1+\delta}, \;\; \mbox{for $\eta\neq 1$},
 \nonumber   \\
 \xi(E) &\sim& |\log E|^{-2+\delta}, \;\; \mbox{for $\eta=1$}.
 \label{xiE}
 \end{eqnarray}
In the case of white-noise disorder,
\begin{equation}
\label{eq:whitecor}
 [ m(x) m(y) ]_{\rm ens} = g \; \delta(x-y),
\end{equation}
(where $g$ is the constant which controls the strength of randomness) 
the values of $\eta$ and $\delta$ are $1$ and $3$, respectively.

 Let us briefly explain the numerical methods to calculate the DOS and
 LL\cite{KI}.
  We set the random mass $m(x)$ in Eq.(\ref{Hdirac}) as multi-kink
  configuration and vary the height of kinks randomly (see Fig.1). 
 Our choice of the random-mass configurations simplifies the problem to solve 
 the Dirac equation for Eq.(\ref{Hdirac}) and we can calculate the energy 
 eigenfunctions systematically using {\it transfer-matrix} methods. 
We also introduce {\it an imaginary vector potential} into
 this system. By varying the magnitude of vector potential,
 we can obtain the localization length of the states 
 {\em without} imaginary vector potential as 
 a function of energy and the correlation length of the random mass.
This idea is based on the work by Hatano and Nelson\cite{HN}.
Localized state has the wave function like,
\begin{equation}
\Psi_0(x)\simeq \exp \Big(-{|x-x_c| \over \xi_0}\Big),
\label{Psi_0}
\end{equation}
where $\xi_0$ is the localization length.
Wave function in the presence of an imaginary vector potential $iA$
is obtained as follows by the ``imaginary" gauge transformation,
\begin{equation}
\Psi_A(x)\simeq \exp \Big(-{|x-x_c| \over \xi_0}-A(x-x_c)\Big).
\end{equation}
Then if $A$ exceeds inverse of the localization length of the original wave
function $\Psi_0$, $\Psi_A$ becomes unnormalizable.
By increasing the value of $A$ and searching normalizable wave function
with the same energy, we can obtain the value of $\xi_0$.

 First we show the energy dependence of the DOS and LL
 for the white-noise disorder
 for comparison with the power-law decaying cases (Fig.3).
We fit the energy dependence of the DOS in the log-log plot by the
 following function\footnote{Here we set the constant $g$ unity in the term
 $|\log E/2g|$ for the convenience of the fitting.},
 \begin{equation}
 f(x) = A + B x + C \log(x) \;\;\;\; (x=|\log E|),
 \label{fitfunction}
 \end{equation}
 where $B$ and $C$ are parameters which correspond to the exponents
 $\eta$ and $\delta$ in Eq.(\ref{rhoE}), respectively. 
The result is shown in the figure of the DOS data, and
$B$ and $C$ are estimated as $B=1.25$ and $C=3.1$ (see Fig.3).  
>From the analytical result, we have $\eta=1$ and $\delta=3$, and therefore
the numerical calculations are in good agreement with the analytical ones.

Let us turn to the LL.
 From the linearity of data in the log-linear plot (top left figure),
 we expect that $\xi(E)$ for the white-noise disorder is parametrized
as follows,
 \begin{equation}
 \xi(E) = F|\log E/2g| = F|\log E| + G,
 \label{LLwhite}
 \end{equation}
 where $g$ is defined in Eq.(\ref{eq:whitecor}), and
$F$ and $G$ are constants. The
 constant $g$ is estimated as $0.90$ from the linear
 fitting, and this is consistent
 with our setting of the numerical calculation in which $g=1$.
We also show the LL data with the log-log plot in Fig.3 (top right). 
>From this, it is obvious that the parametrization in terms of
Eq.(\ref{LLwhite}) gives better fit for the LL than any
parametrization with $\eta\neq 1$ as expected.
This means that our calculation gives reliable results.

\begin{figure}
\begin{center}
\unitlength=1cm
\begin{picture}(12,4)
\put(-4.5,1){
\centerline{
\epsfysize=4cm
\epsfbox{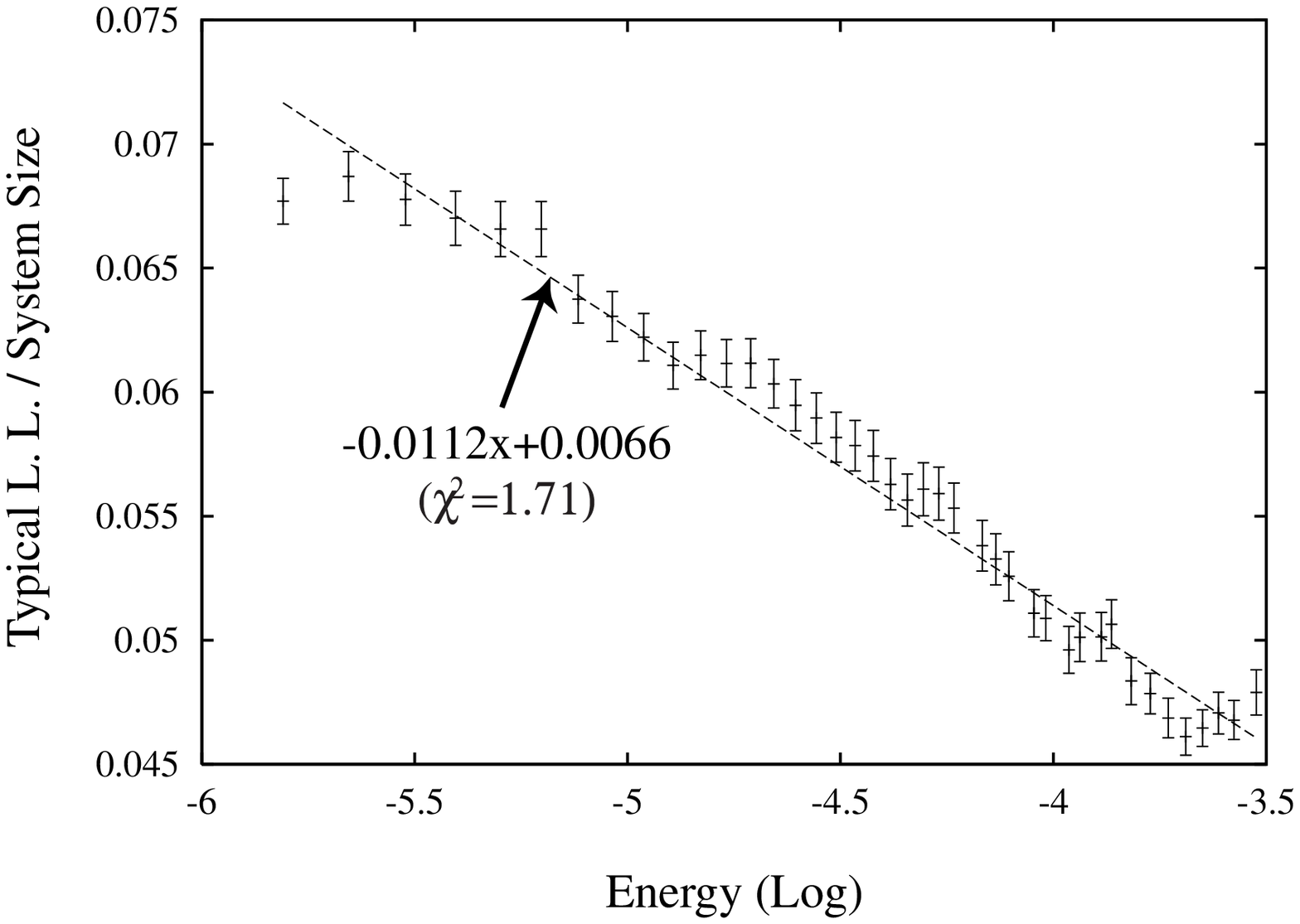}
}}
\put(1.5,1){
\centerline{
\epsfysize=4cm
\epsfbox{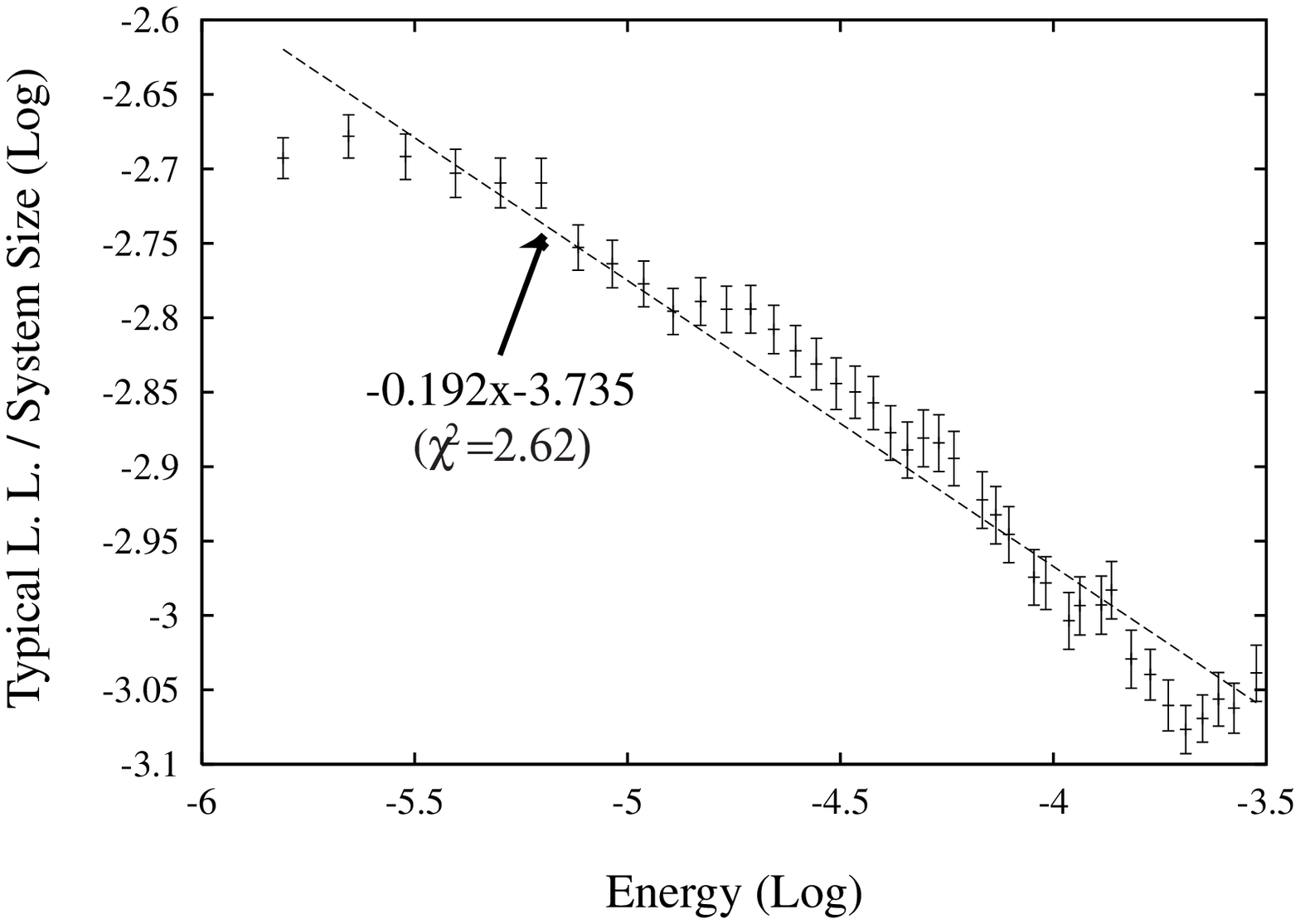}
}}
\put(-4.5,-3.5){
\centerline{
\epsfysize=4cm
\epsfbox{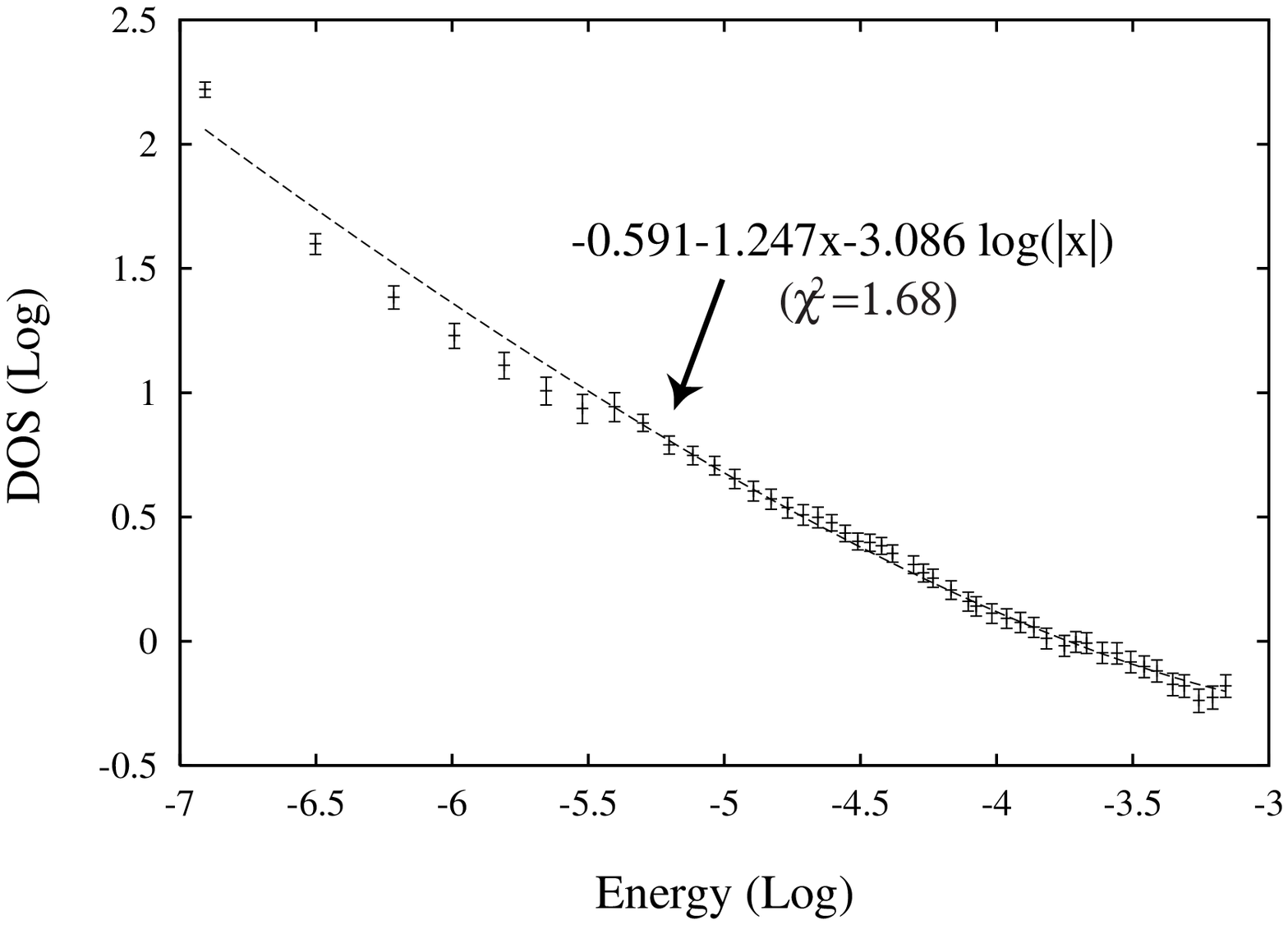}
}}
\label{fig:lldoswhite}
\end{picture}
\vspace{32mm}
\caption{\small Energy dependence of the LL and DOS in the
 system with white-noise disorder:
 We set L(system size)=102.4 and 1024 kinks in these systems. The
 constant $g$ in Eq.(\ref{eq:whitecor}) is unity in this calculation.
 The lines in the top two figures are the results of linear fitting.
 The result of non-linear fitting for DOS is shown as the curve in the bottom figure.}
\end{center}
\end{figure}
\vspace{5mm}

 In the previous papers\cite{KI}, we found that the value of
the parameter $\eta$ in Eq.(\ref{rhoE}) is about 0.5 for the power-law
 correlated disorders. 
 We also showed that the the value of $\eta$ is rather insensitive to
 the power-law exponent of correlation, $\gamma$ in Eq.(\ref{ffm1}). 
 Since $\eta=1$ for the white-noise disorder, we concluded that
 there is a ``phase transition"
 as the correlation of the random variables is varied from the short-range to
 long-range correlations.
 However in Ref.\cite{KI}, we used the {\em old-fashioned} FFM 
 for generating random correlated numbers and therefore we suspect that
 long-range behavior of correlated random numbers was not produced correctly. 
 In this paper we calculate energy dependence of the LL
 and DOS for the power-law correlated disorder by
 using {\it modified} FFM in order to examine our previous results.

 Numerical results of the LL and DOS obtained by
 using MFFM are given in Fig.4 for $\gamma=0.4$ and $\gamma=0.2$.  
We show the LL data in log-linear and also log-log plots
as in the previous white-noise case.
In the both cases of $\gamma=0.2$ and $\gamma=0.4$, the results in 
Fig.4 show that $\eta \sim 0.5$ gives better fit than that of the
white-noise parameters as we found in the previous paper\cite{KI}.
Therefore we have confirmed the validity of the previous result, i.e,.
the LL and DOS exhibit different behaviors as the correlation of the
random variables is changed from the short-range to long-range ones.
We hope that this ``phase transition" is observed in future by experiments
of the random spin chains.

\begin{figure}
\begin{center}
\unitlength=1cm
\begin{picture}(12,4)
\put(-4.5,3){
\centerline{
\epsfysize=3.6cm
\epsfbox{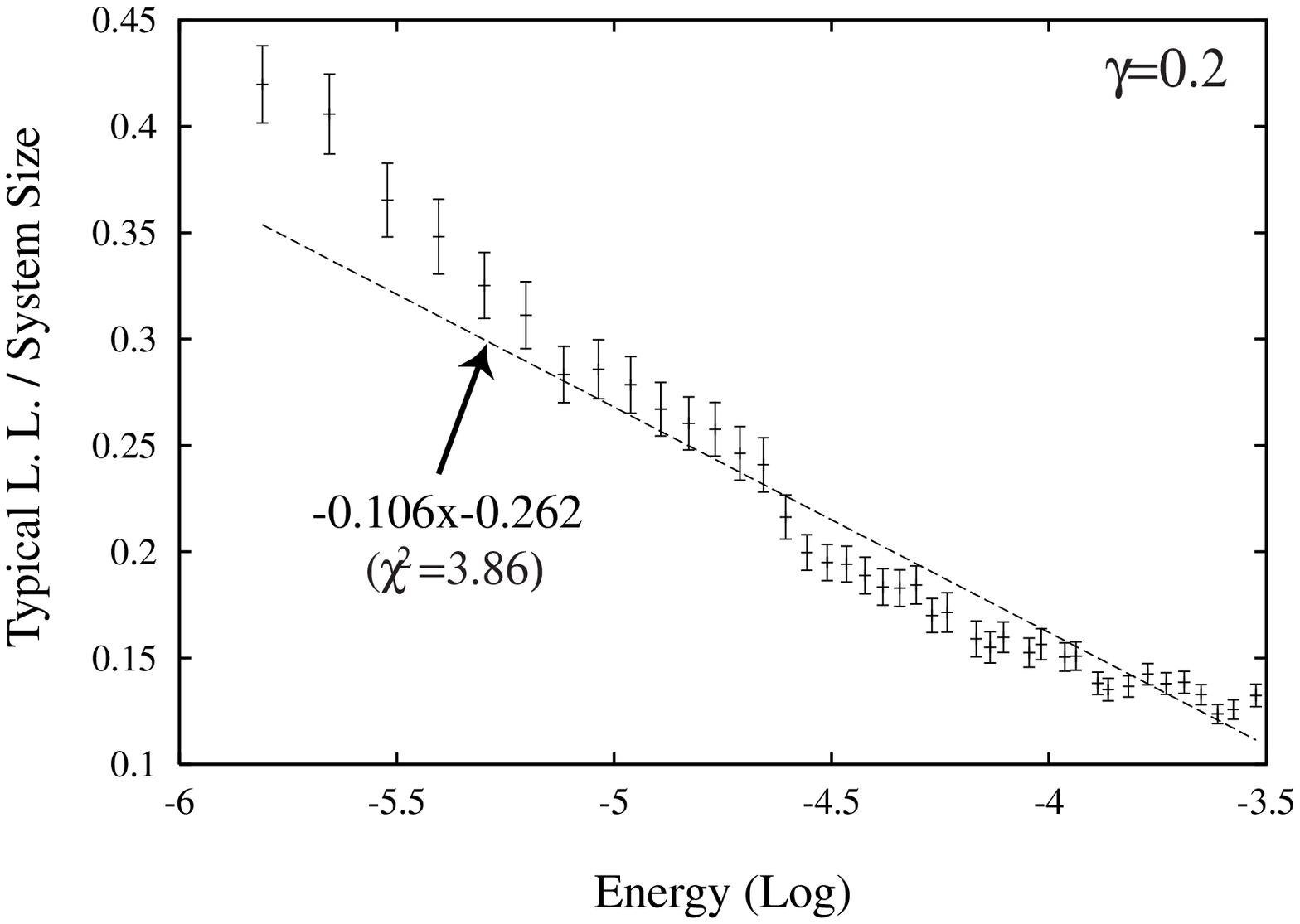}
}}
\put(1,3){
\centerline{
\epsfysize=3.6cm
\epsfbox{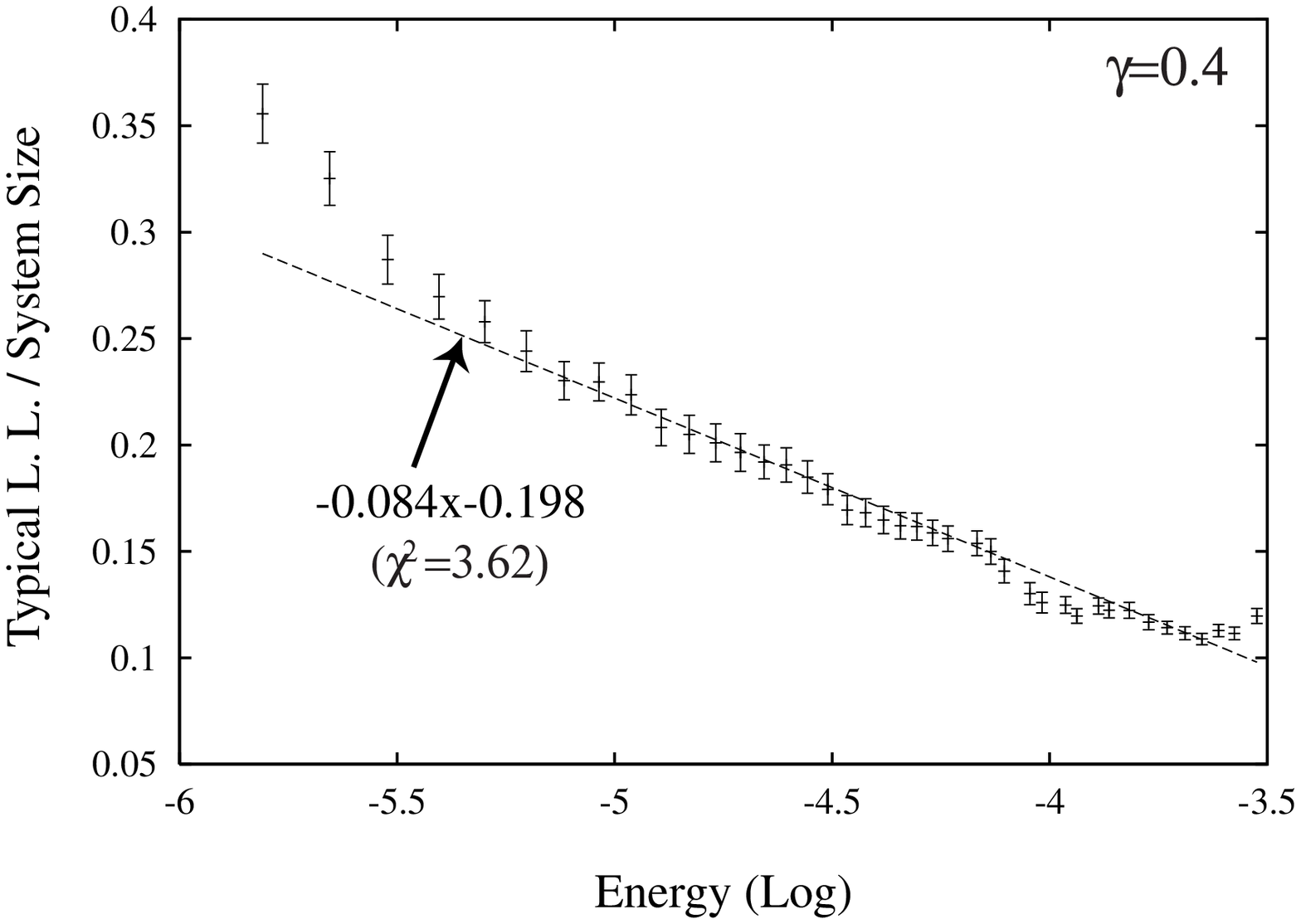}
}}
\put(-4.5,-1){
\centerline{
\epsfysize=3.6cm
\epsfbox{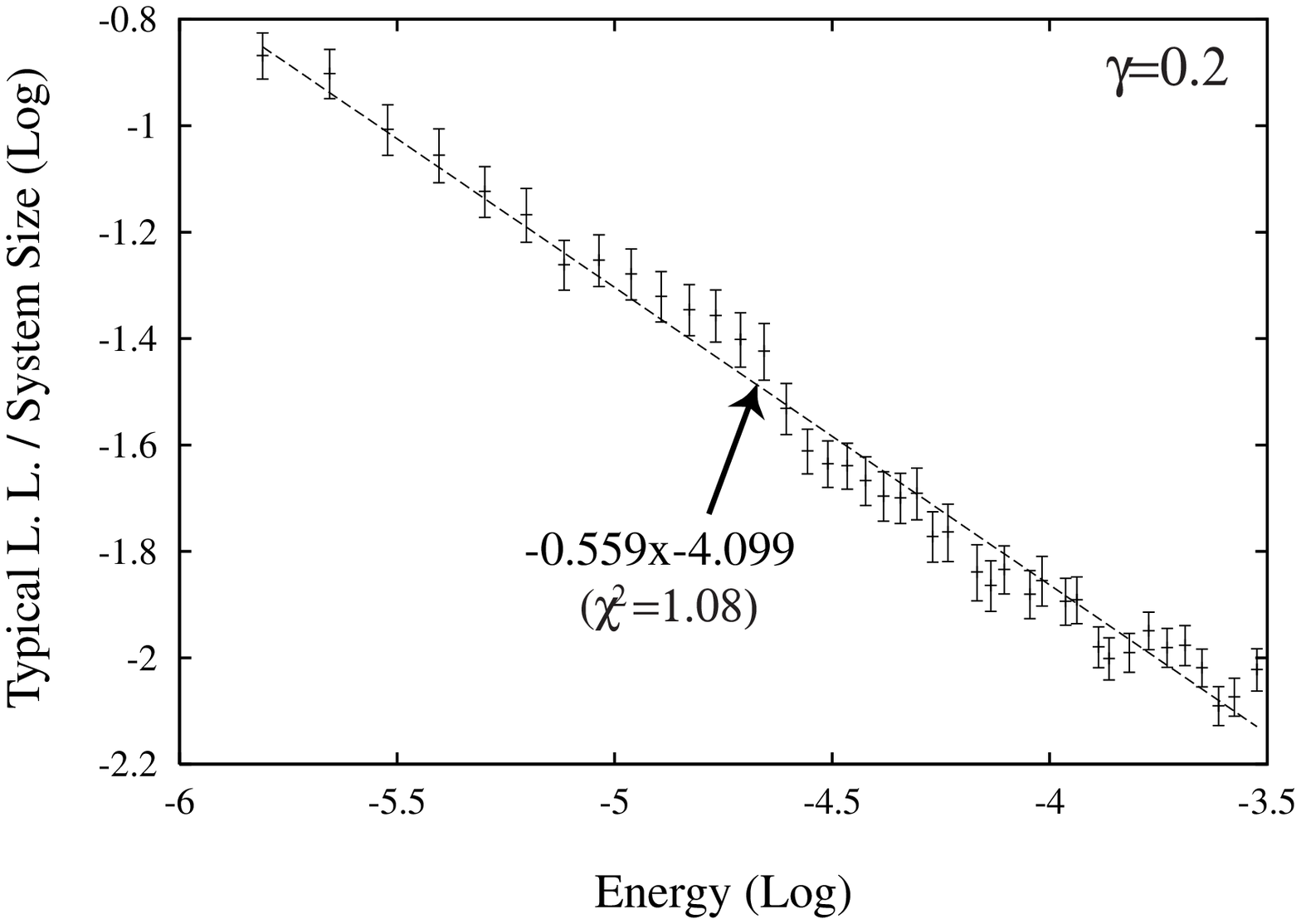}
}}
\put(1,-1){
\centerline{
\epsfysize=3.6cm
\epsfbox{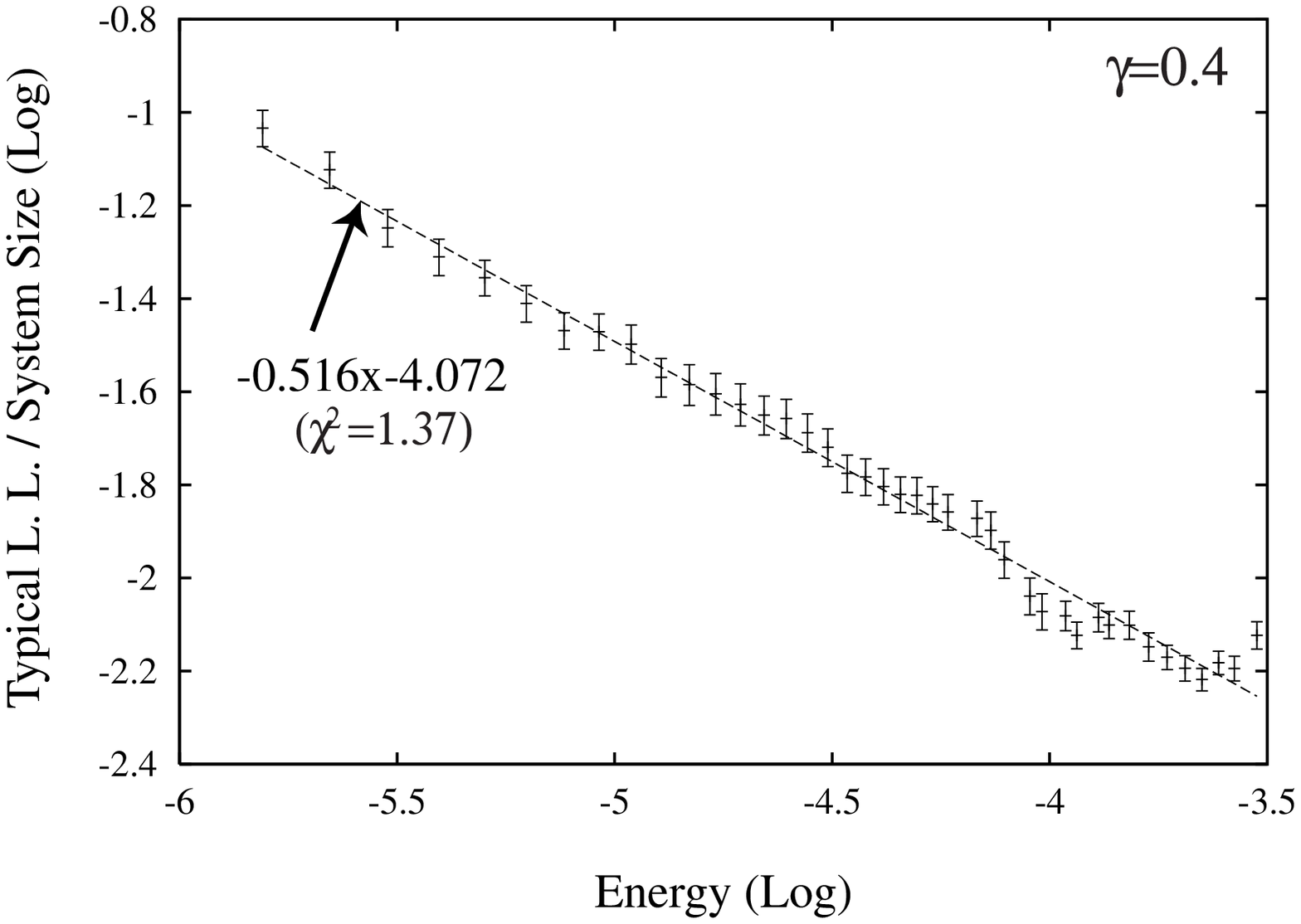}
}}
\put(-4.5,-5){
\centerline{
\epsfysize=3.6cm
\epsfbox{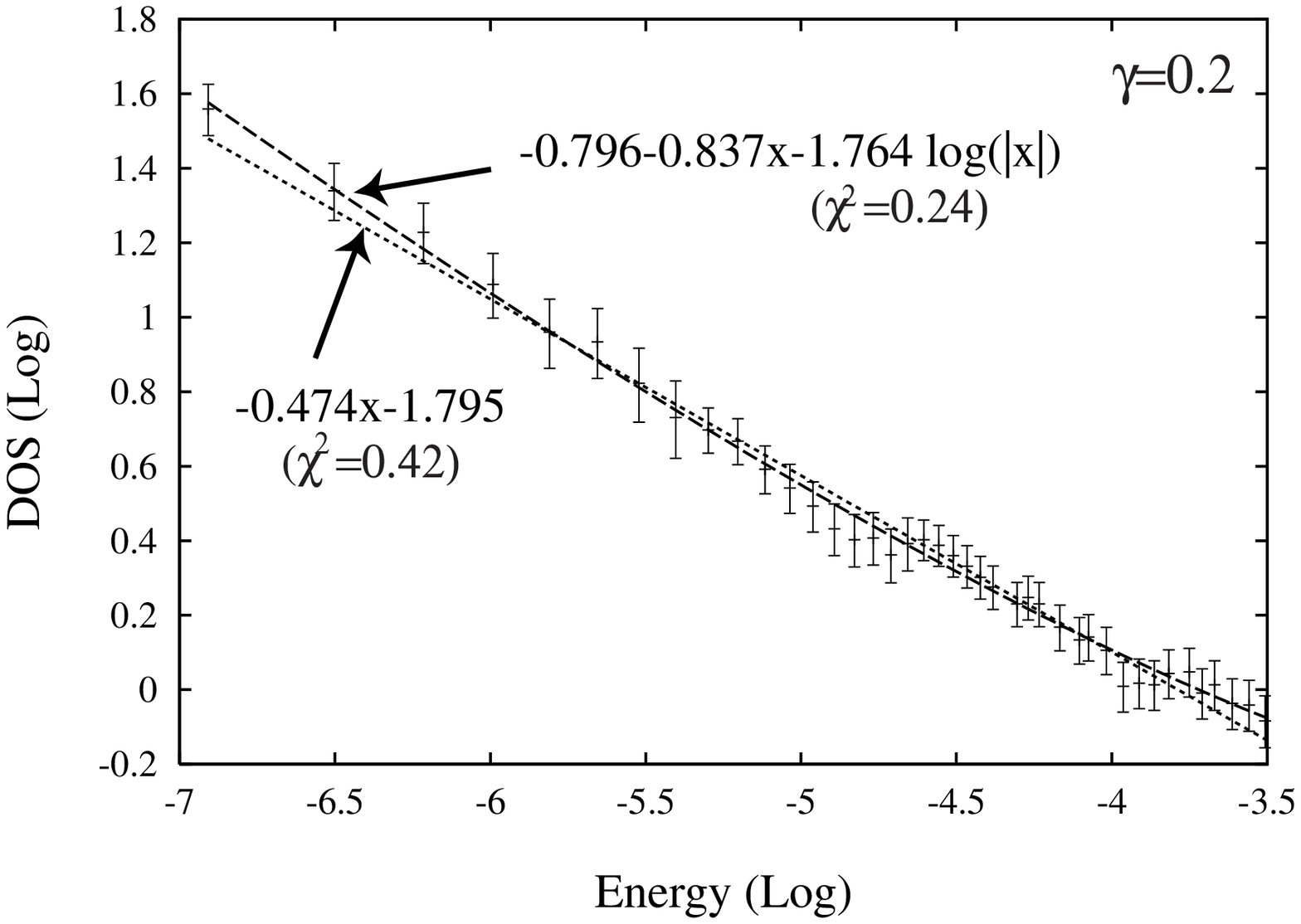}
}}
\put(1,-5){
\centerline{
\epsfysize=3.6cm
\epsfbox{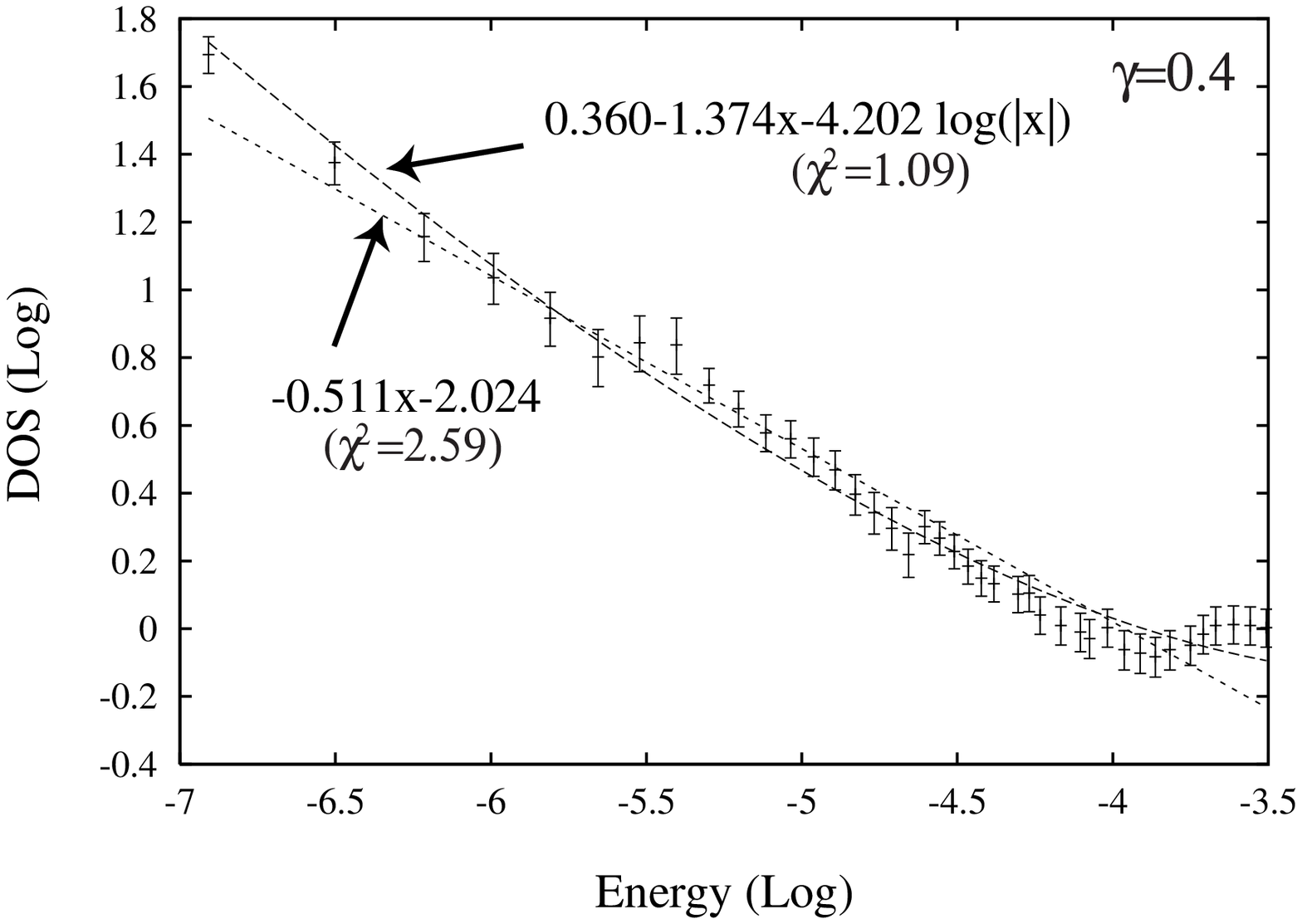}
}}
\label{fig:lldos}
\end{picture}
\vspace{47mm}
\caption{\small Energy dependence of the LL and DOS in the
 system with power-law correlated disorders:
 We set L(system size)=102.4 and 1024 kinks in these systems. 
 $\gamma$'s in the figures are defined in Eq.(\ref{ffm1}). 
 We show the energy dependences of LL in log-linear and log-log plots.
 We also show the energy dependence of DOS in log-log plot.
 The fitting results are shown as the lines and the curves in each
 figure (see also the text).}
\end{center}
\end{figure}
\vspace{5mm}


\section{Zero-energy wave functions and multifractality}
In this section we study the multifractality of zero-energy 
wave functions in
 the system with nonlocally correlated disorders.
Behavior of the wave functions must give us more detailed information
on the effects of the nonlocal correlations of the random disorders.
In the case of the {\em white-noise} disorder,
the zero-energy wave functions are exactly obtained 
 and it is not so difficult to calculate the ensemble-averaged
 correlation of them\cite{BF,Shelton}
 \begin{equation}
 W_q(l,L)=[ |\psi^{\dagger}(x)\psi(x+l)|^q ]_{{\rm ens}},
 \label{Wq}
 \end{equation}
where $L$ is the system size, $q$ is a number and $\psi(x)$ is the
zero-energy wave function.
We shall calculate $W_q(l,L)$ numerically for various nonlocally correlated
disorders.
It is easy to see that the above $W_2(l,L)$ is related with
the spin-spin correlation function,
\begin{equation}
[\langle S^z(x)S^z(y) \rangle ]_{\rm ens} = 
[\langle :\psi^\dagger\psi(x): :\psi^\dagger\psi(y): \rangle]_{\rm ens},
\label{SSC}
\end{equation}
where the normal ordering $:{\cal O}:$ is taken with respect to the
half-filled state.
The above spin-spin correlation function is dominated by 
the nodeless lowest-energy
mode for each configuration of the random mass.
For the white-noise correlation, relationship between the averaged Green 
function and the lowest-energy wave functions is discussed rather 
in detail in Ref.\cite{BF}.
By using the transfer-matrix methods, it is not so difficult to
obtain the wave function of the lowest energy and to calculate the
correlation (\ref{Wq}) numerically.
In the non-random Heisenberg spin chain with a uniform exchange 
coupling $J_n=J$
for all $n$, the above correlator decays as 
\begin{equation}
[\langle S^z(x)S^z(y) \rangle]_{\rm ens} \propto {1\over |x-y|},
\label{uniformco}
\end{equation}
whereas in the spin-Peierls state with $J_n=J+(-)^nm_0$,
\begin{equation}
[ \langle S^z(x)S^z(y)\rangle ]_{\rm ens} \propto \exp (-m_0|x-y|).
\label{dimerco}
\end{equation}
Furthermore for the random exchange of the white-noise correlation,
which corresponds to the limit $\lambda \rightarrow 0$ of Eq.(\ref{expC}),
the correlator $W_q(l,L)$ was calculated both analytically\cite{BF} and 
numerically\cite{TTIK},
\begin{equation}
[ |\psi^\dagger\psi(x)\psi^\dagger\psi(y)|^q ]_{\rm ens}
\propto {1\over |x-y|^{3 \over 2}}.
\label{wnco}
\end{equation}
$W_q(l,L)$ essentially measures the spatial correlation of the first
and second peaks of the wave functions.
On the other hand, the spin-spin correlator behaves as 
\begin{equation}
[\langle S^z(x)S^z(y)\rangle ]_{\rm ens}
\propto {1\over |x-y|^{2}}
\end{equation}
Relation between the spin-spin correlation function
$[\langle S^z(x)S^z(y)\rangle ]_{\rm ens}$
and $W_q(l,L)$ is discussed in Ref.\cite{SFG}.
Therefore if the long-range behavior of $W_q(l,L)$ changes 
as a result of the long-range correlation of the random variables,
then we can expect that the spin correlation functions also
behave differently from those of the white-noise case.

We must notice one technical difference between analytical and numerical
studies. 
Normalization of the zero-energy wavefunctions plays an 
important role in the numerical study because they are
genuinely normalizable only for specific $m(x)$.
More explicitly, in solving the Dirac equation  
we imposed the periodic boundary condition on the system.
In contrast,
boundary condition is not imposed on the wave function in the analytical 
calculation\cite{SFG,BF,Shelton}. 

It is expected that the correlation functions of zero-energy wavefunction
exhibit multi-fractal scaling,
\begin{equation}
W_q(l,L) \sim L^{-1-\tau(q)}|l|^{-y(q)}.
\label{multiF}
\end{equation}
For the system of the white-noise random mass $[m(x)m(y)]_{\rm ens}
\propto \delta(x-y)$, 
$W_q(l,L)$ is obtained as follows 
for $L \rightarrow \infty$ by the analytical calculation\cite{BF},
\begin{equation}
W_q(l,L) \sim \frac{\tilde{W}(q^2l)}{L}
\label{Wq2}
\end{equation}
where
\begin{equation}
\tilde{W}(l)=\int^{\infty}_0dk \frac{k^2}{(1+k^2)^4}e^{-lk^2}.
\label{Wq3}
\end{equation} 
For large $l$ 
\begin{equation}
\label{Wq4}
W_q(l,L) \sim 1/(l^{3/2}L)
\end{equation}
and therefore it is expected that $\tau(q)$ and $y(q)$ are independent of 
$q$ and $\tau(q)=0$ and $y(q)=3/2$ for the
white-noise disorder.
In this paper,
we shall numerically obtain the zero-energy wave functions and
calculate $y(q)$ under various types of disorder using the MFFM.
Typical example of the zero-energy wave function obtained numerically is
given in Fig.5.
\begin{figure}
\begin{center}
\unitlength=1cm
\begin{picture}(17,4.5)
\centerline{
\epsfysize=4cm
\epsfbox{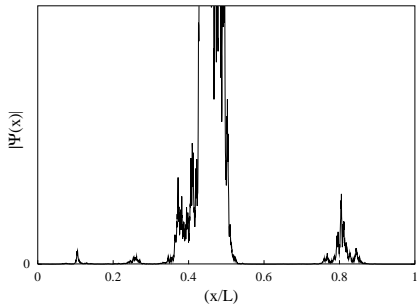}
}
\label{fig:white}
\end{picture}
\vspace{-10mm}
\caption{\small An example of zero-energy wavefunction under
 white-noise disorder.}
\end{center}
\vspace{0mm}
\end{figure}

First we numerically calculate $W_q(l,L)$ in the case of the white-noise 
disorder.
Values of $y(1)$ and $y(2)$ are obtained from the calculations of
$W_q(l,L)$ as a function of $l/L$. (See Fig.6.)
The value of $y(2)$ is related with the power of algebraically decaying 
$[ S^z(x)S^z(y) ]_{\rm ens}$ in the {\em crossover region}\cite{SFG}.
>From Fig.6, the values of $y(1)$ and $y(2)$ for white-noise disorder are
both about $1.5$ and this result is in agreement with Eq.(\ref{Wq4}). 
Similarly we also calculate  $y(1)$ and $y(2)$ for exponentially-correlated 
disorder.
Numerical calculations are shown in Fig.6 ($\lambda=0.5$
and $25$) and the results are summarized in Table 1. 
For small values of $\lambda/L$, $y(2)$ is also about $1.5$. On the other
hand for $\lambda=25$, $y(2) \sim 2.2$. This reflects
the effect of the nonlocal correlations.
The above result suggests that nontrivial change occurs in the behavior
of $W_q(l,L)$ for the power-decay correlated mass 
$[ m(x)m(y) ]_{\rm ens}\propto {1\over |x-y|^\gamma}$.

\begin{figure}
\begin{center}
\unitlength=1cm
\begin{picture}(12,3.5)
\put(-1.5,1){
\centerline{
\epsfysize=4cm
\epsfbox{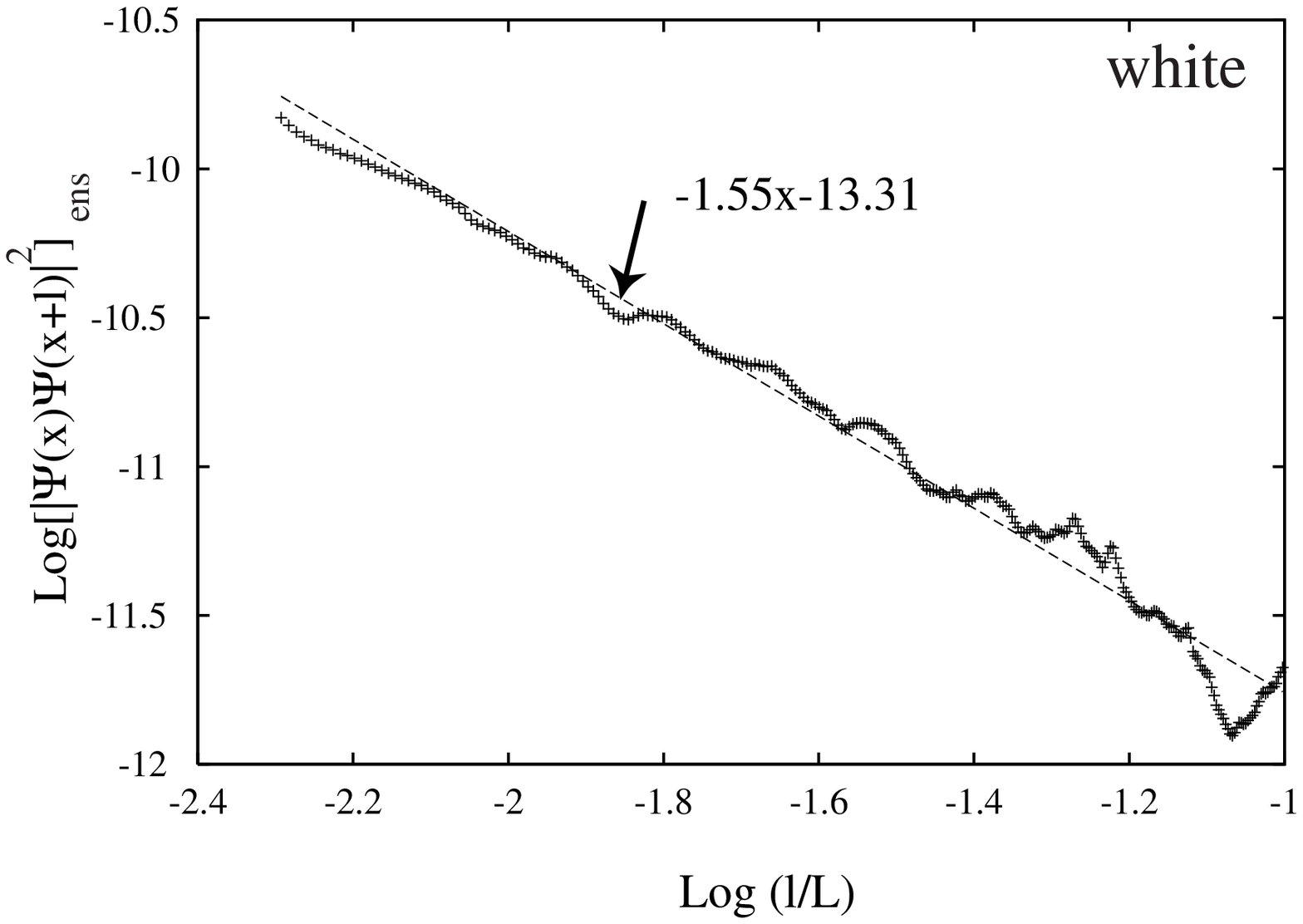}
}}
\put(-4.5,-3.5){
\centerline{
\epsfysize=4cm
\epsfbox{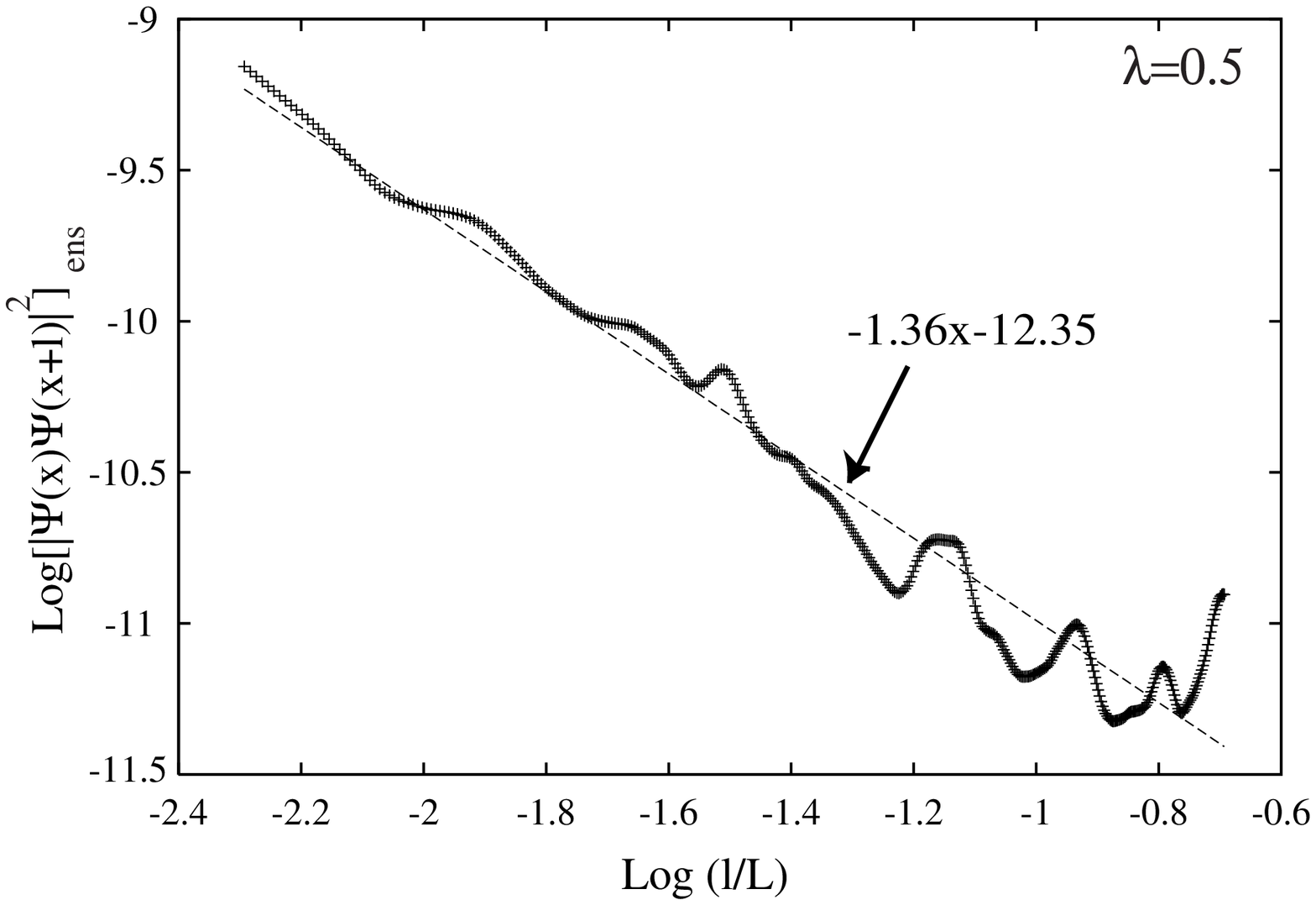}
}}
\put(1.5,-3.5){
\centerline{
\epsfysize=4cm
\epsfbox{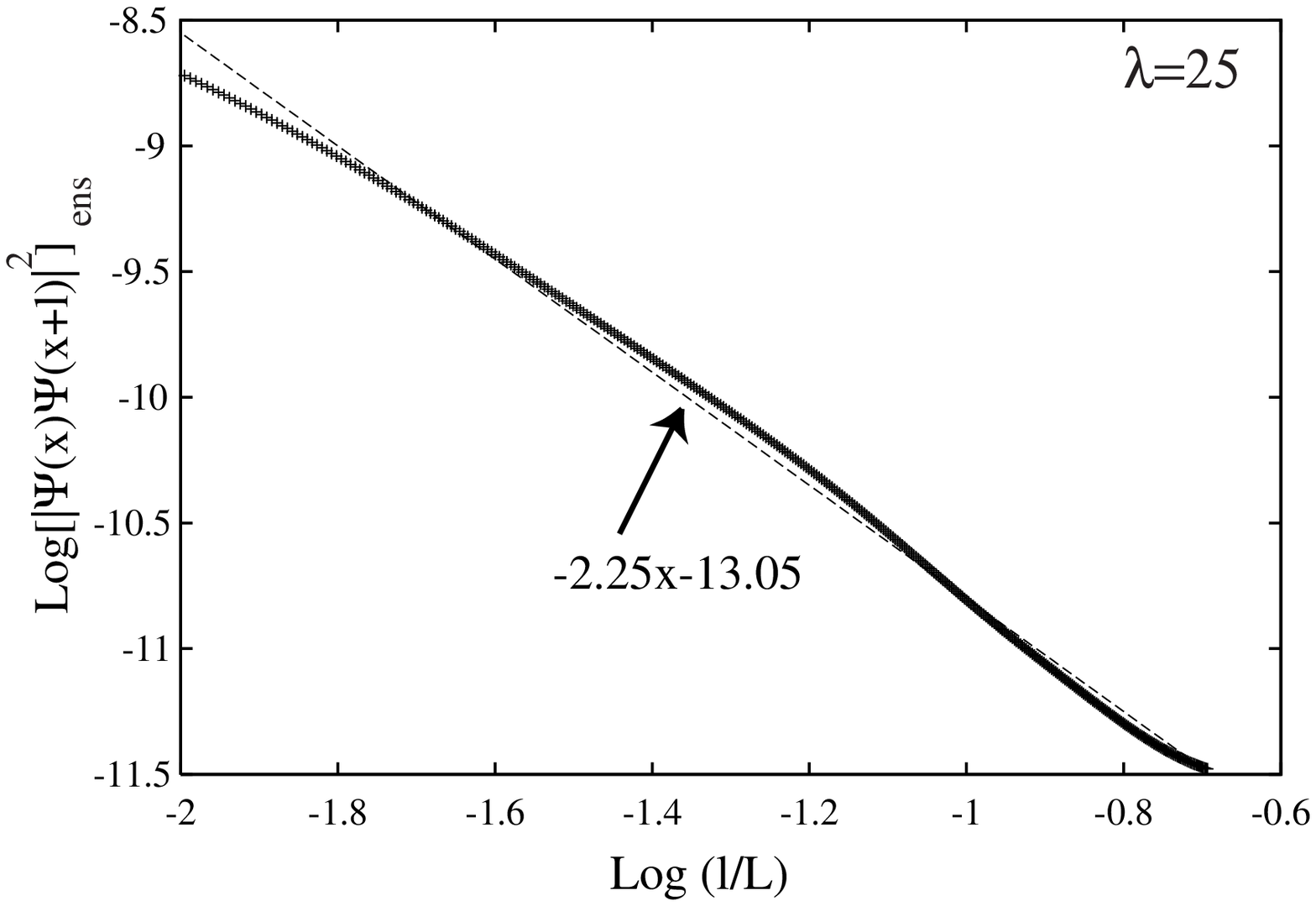}
}}
\label{fig:whitecor}
\end{picture}
\vspace{3.5cm}
\caption{\small The correlation function $W_q(l,L)$ as a function of $l/L$
for the white-noise disorder 
 $[ m(x)m(y) ]_{\rm ens} = \delta (x-y)$ and exponentially correlated
 disorder $[ m(x)m(y) ]_{\rm ens} = \frac{1}{2\lambda}
 \exp(-\frac{|x-y|}{\lambda})$:
 We set $L=102.4$ and number of sites $2048$ here.}
\end{center}
\end{figure}

\begin{center}
 \begin{tabular}{c|ccccc}
  $\lambda$ & $0$(white) & $0.05$ & $0.5$ & $5$ & $25$ \\ \hline
  $y(1)$ & $1.19$ & $1.05$ & $1.01$ & $1.27$ & $1.25$ \\
  $y(2)$ & $1.55$ & $1.46$ & $1.36$ & $1.62$ & $2.25$
  \end{tabular}
 \end{center}
\vspace{5mm}
 {Table 1: \small {Fractal exponent $y(1)$ and $y(2)$:
 The parameters are set as follows; $[ m(x)m(y) ]_{\rm ens} = \exp (
 |x-y|/\lambda) $,
 $L=102.4$ and $2048$ kinks. The range of the data is $-2.3<\log
 (l/L)<-0.7$.}}
 
\vspace{5mm}

Next we calculate  $y(1)$ and $y(2)$ for the case of the long-range correlated
disorders with power-law decay.
We show the results in Table 2. (Examples of correlation are shown
in Fig.7.)
The values of $y(1)$ are about ${3\over 2}$ in almost all cases
except $[ m(x)m(y) ]_{\rm ens} = 0.01/|x-y|^\gamma$.
This result may indicate universal properties of the single-particle
Green's function.
However in all cases, $y(1)\neq y(2)$.
This is in sharp contrast to the case of the white-noise disorder.
In each cases, the value of $y(2)$ is larger than
${3\over 2}$. 
More precisely, the estimated value $y(2)$ is larger for smaller 
value of $\gamma$ and some of the $y(2)$ exceeds 2.
On the other hand for large $\gamma$, $y(2)$ is getting close to
${3\over 2}$. 
This reflects the fact that the effect of correlation becomes
larger for smaller value of $\gamma$.
The long-range correlation of the random mass {\em suppresses} the appearance
of a large variety of wave functions.
However our numerical study indicates that the system is still in {\em a 
critical region} with the power-decaying correlators.

\begin{figure}
\begin{center}
\unitlength=1cm
\begin{picture}(12,4.7)
\put(-4.5,1){
\centerline{
\epsfysize=4.5cm
\epsfbox{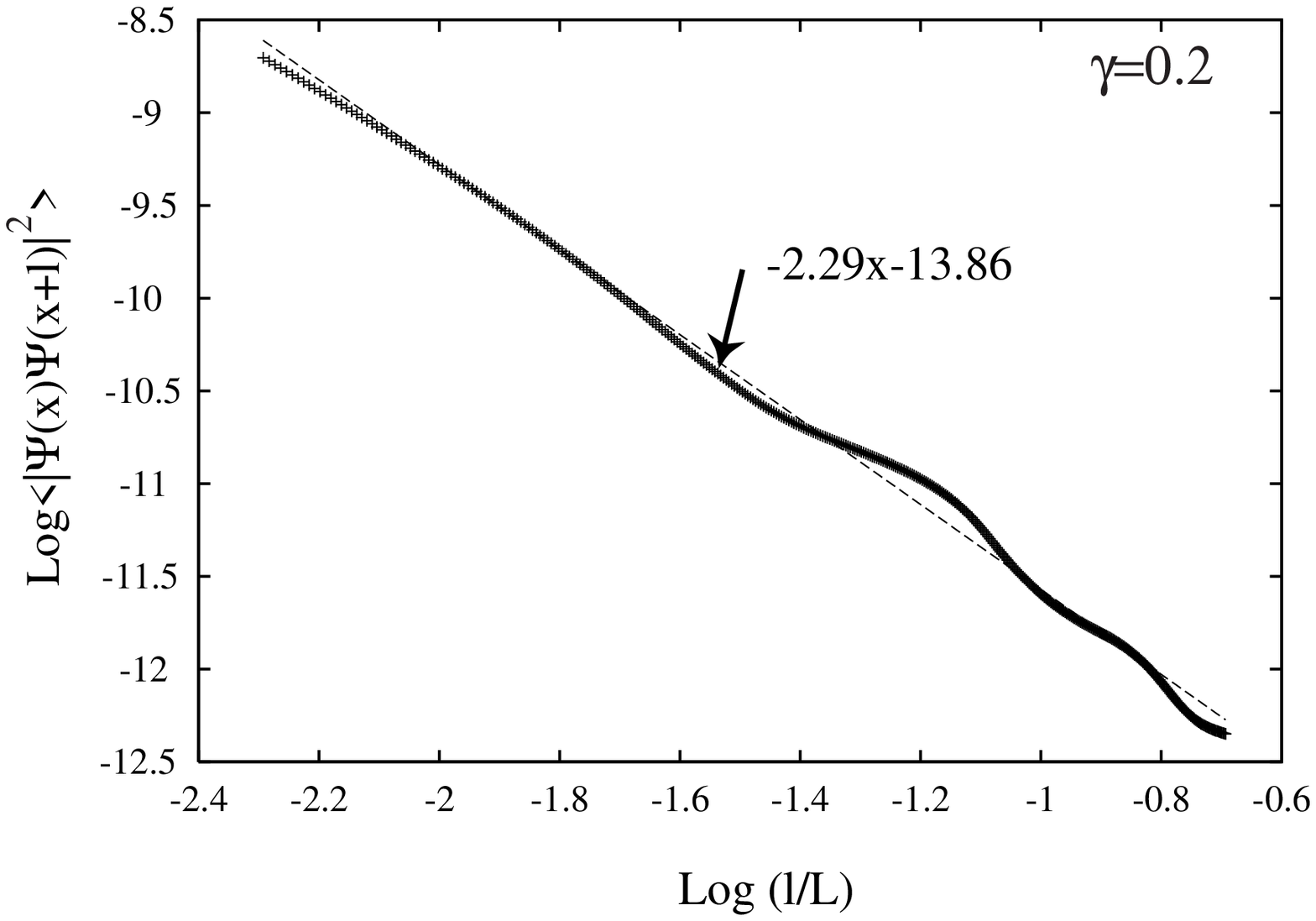}
}}
\put(2,1){
\centerline{
\epsfysize=4.5cm
\epsfbox{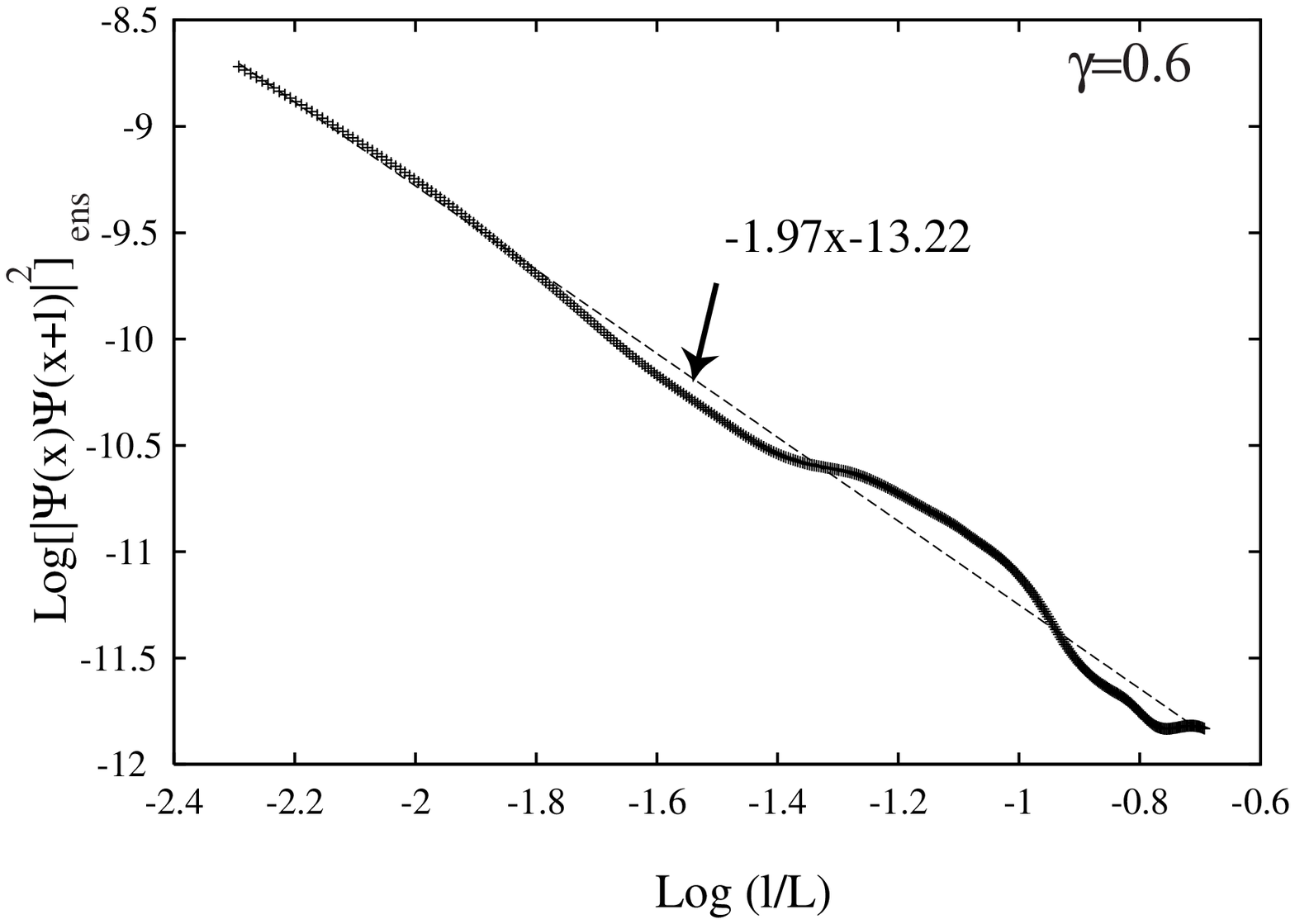}
}}
\label{fig:powercor}
\end{picture}
\vspace{-0.7cm}
\caption{\small Examples of correlation function $W_q(l,L)$ under the power-law
 correlated disorder $[ m(x)m(y) ]_{\rm ens} = C / |x-y|^{\gamma}$:
 We set L(system size)=102.4, 2048 kinks and C=0.1 here. }
\end{center}
\end{figure}

\begin{center}
 \begin{tabular}{c|cccc|c|c|c|c} 
   $$ & $$ & \# of & $$ & range of & \multicolumn{4}{c}{$y(1) \ |
  \ y(2)$} \\ \cline{6-9}
  $$ & \multicolumn{1}{c}{\raisebox{2.5ex}[0pt]{$L$(size)}}
  & \raisebox{1.0ex}[0pt]{kinks}&
  \multicolumn{1}{c}{\raisebox{2.5ex}[0pt]{$C$}} &
    \raisebox{1.0ex}[0pt]{data $\log (l/L)$} & $\gamma=0.2$
  & $\gamma=0.4$ & $\gamma=0.6$ & $\gamma=0.8$ \\ \hline
  I & $51.2$ & $1024$ & $1$ & $-2.3\sim-0.7$ & $1.37\ | \ 2.39$ & $1.47 \ | \ 2.41$ & $1.45\ | \ 2.20$ & $1.27\ | \ 2.03$ \\
  II & $102.4$ & $2048$ & $0.1$ & $-2.5\sim-2$ & $1.70\ | \ 2.29$ &
  $1.60 \ | \ 2.13$ & $1.46\ | \ 1.97$ & $1.33\ | \ 1.85$ \\
  III & $204.8$ & $4096$ & $0.01$ & $-2.3\sim-1.5$ & $1.41\ | \ 2.27$ &
  $1.26\ | \ 2.03$ &  $1.12\ | \ 1.85$ & $0.97\ | \ 1.70$ 
 \end{tabular}
 \end{center}
\vspace{5mm}
{Table 2: {\small Fractal exponent $y(1)$ and $y(2)$:
 The parameters are shown in the table. $C$ and $\gamma$ are defined as
 $[m(x)m(y)]_{\rm ens} = C / |x-y|^{\gamma}$.}}
\vspace{5mm}

\section{Conclusion and Discussion}
In this paper, we studied $S={1\over 2}$ spin chain with nonlocally-correlated
random exchange coupling via its low-energy effective field theory, 
random-mass Dirac fermions.
In particular we investigated the effects of the long-range correlation
of the random variables.
In order to generate random numbers with power-decaying correlations, 
we employed the MFFM.
We calculated the DOS, LL and correlators of the zero-energy functions
$W_q(l,L)$, which give important information about low-energy states
of the system.
The results indicate that there appear nontrivial effects of the
long-range correlation in various physical quantities.
Energy dependence of the DOS and LL and the long-distance
behavior of $W_q(l,L)$ are changed by the correlations.
As $W_q(l,L)$ is closely related with the spin correlation 
function $[\langle S^z(x)S^z(y) \rangle]_{\rm ens}$, the above result
indicates that the spin correlators also exhibit different
long-distance behavior from that of the short-range
correlated random variables.

Let us briefly discuss qualitative nature of the ``new phase"
which appears as a result of the long-range correlation of
the random variables.
In the short-range white-noise case, the DOS behaves as $\rho(E)\sim E^{-1}
|\log E|^{-3}$ and the LL $\xi(E)\sim |\log E|$.
On the other hand for the long-range correlated case,
$\rho(E)\sim \xi(E) \sim E^{-{1\over 2}}$.
This means that the low-energy states become more extended
by the effect of
the long-range correlations whereas the low-energy DOS becomes smaller
because of it through the Thouless formula.
As the localization length $\xi_0$ parametrizes the typical
wave functions as in (\ref{Psi_0}), it measures the width
of the largest peak of the wave functions.
Furthermore we found that $W_q(l,L)$, in particular $q=2$,
tends to decay more rapidly by the long-range correlations.
As $W_q(l,L)$ measures the correlations between the first and second (or third)
largest peaks of the wave functions, this result indicates that 
the second largest peak 
becomes smaller compared to the first one and/or the distances between the 
first and second largest peaks tend smaller by the long-range correlations.
In the picture of the RS phase of the spin chains, the above results
imply that amplitude of the spin-singlet pair becomes a smooth 
decreasing function
of the distance between spins in the pair as a result of the long-range
correlations of the exchange couplings.
In other words, the ``randomness"
or ``variety of configurations"  of the RS phase is suppressed by 
the long-range correlations.
Actually in the Dirac fermions with the mass of the single-kink
configuration like $m(x)\propto \theta(x)$,
where $\theta(x)$ is the step function, an exactly massless mode 
appears in the vicinity of the kink of $m(x)$
where $m(x)$ changes its sign\cite{NS}.
This localized massless mode corresponds to an unpaired spin in the dimer state
of the spin chains which appears as a result of the sign change of $m(n)$
in (\ref{Jn}).
In the random-mass Dirac fermions, the random mass $m(x)$ vanishes
at various points $x=x_i$ $(i=\mbox{an integer})$ and roughly speaking 
low-energy modes are given
by linear combinations of the localized modes in the vicinity of the
kinks of $m(x)$\cite{TTIK} like $\sum_i C_i|i\rangle$
where $|i\rangle$ is the localized state at $x_i$ and $C_i$'s
are coefficients.
The long-range correlation of the random mass keeps distance
between adjacent kinks $|x_i-x_{i+1}|$
longer and as a result it makes the LL
of the low-energy states larger.
On the other hand, mixing amplitude of the modes localized
in the vicinity of different kinks of $m(x)$ tends smaller by the
smooth behavior of $m(x)$, in other words
some coefficient $C_{i_0}$ dominates the others $C_{i\neq i_0}$
and $C_i$ decreases smoothly as $|x_{i_0}-x_i|$ increases.
This consideration indicates that the low-energy DOS tends smaller
by the long-range correlation as we observed by the numerical calculation.
In the new phase, most of spins make a singlet pair with their
nearest-neighbor spin. 
Spins in kinks of $m(n)$ embedded in the dimer state have weak
antiferromagnetic correlations
with each other and generate low-energy spin excitations.



\begin{thebibliography}{99}
\bibitem{SSCF}The spin-spin correlation function of the random exchange
Heisenberg spin chain is obtained as $[\langle S(x)\cdot S(y)\rangle]_{\rm ens}
={1\over |x-y|^2}$ for the $\delta$-function white-noise 
disorder \cite{DF,SC}. 
Relation between this behavior and $W_q(l,L)$ is 
discussed in Ref.\cite{SFG}.
\end{thebibliography}
\end{document}